\documentstyle[12pt,epsf]{article}
\def\bbox#1{\mbox{\boldmath $#1$}}
\begin{document}
{\bf Preprint FZR-202, November 1997}
\begin{center}

\vspace*{3cm}

{\large \bf Polarization observables in the reaction $NN \to NN \Phi$}\\[6mm]
{\sc  A.I. Titov$^{a,b}$, B. K\"ampfer$^a$, V.V. Shklyar$^b$}\\[3mm]
$^a$Research Center Rossendorf, PF 510119, 01314 Dresden, Germany\\[1mm]
$^b$Bogoliubov Theoretical Laboratory, Joint Institute for Nuclear Research,\\
141980 Dubna, Russia\\[2cm]
{\bf Abstract:}
\end{center}
We study the reaction $NN \to NN \Phi$ slightly above the threshold
within an extended one-boson exchange
model which also accounts for $uud$ knock-out.
It is shown that polarization observables,
like the beam-target asymmetry, are sensible quantities for identifying
a $s \bar s$ admixture in the  nucleon wave function on the
few per cent level.
\newpage
\setcounter{page}{1}
{\bf 1. Introduction:}
The investigation of the ${NN\to NN\Phi}$ reaction
is interesting for several reasons. First, the elementary
total cross section \cite{Sibirtsev,Ko_PL}
is an important input for the calculation of the $\Phi$ production in
heavy-ion collisions \cite{Ko}.
In this case one might expect some change
of the $\Phi$ width \cite{Weise} due to the coupling to the decay
channel $\Phi\to K^+K^-$ and peculiarities according to the
in-medium modification of the kaon properties \cite{Kolomeitsev,Grosse}.
Indeed, such measurements are under way with the $4\pi$ detector FOPI
at SIS in GSI/Darmstadt \cite{Kotte}.
The electromagnetic decay channel $\Phi \to e^+ e^-$ will be investigated
with the spectrometer HADES \cite{HADES} also in GSI.
Note, that a threshold-near measurement
of the total cross section of the reaction $pp\to pp\Phi$ has been performed
at SATURNE \cite{Brenschede} and precision measurements of polarization
observables are envisaged with the ANKE spectrometer
at the cooler synchrotron COSY in J\"ulich \cite{Sapozhnikov}.

Second, the $\Phi$ meson is thought to consist mainly of strange
quarks, i.e. $s\bar s$, with a rather small contribution of the light
$u,d$ quarks. According to the OZI rule \cite{OZI} the $\Phi$ production
should be suppressed if the entrance channel does not possess
a considerable admixture of hidden or open strangeness.
Indeed, the recent experiments on the proton annihilation at rest,
$p\bar p \to \Phi X$ (cf. \cite{Ellis} for a compilation of data), point to
a large apparent violation of the OZI rule,
which is interpreted \cite{Ellis} as a
hint to an intrinsic $s\bar s$ component in the proton.
Analyses of the $\pi N$ sigma term \cite{sigmaterm} also suggest that
the proton might contain a strange quark admixture as large as 20\%.

It would be important and interesting to look for another
clear signal \cite{Ellis,newexp} that might be related directly with the
strangeness content of nucleons. In ref.~\cite{Titov1} it is shown
that the polarization observables in the
$\Phi$ photoproduction are sensitive even to a small strangeness admixture
in the proton. Other investigations concern the polarization
observables in $pp\to pp\Phi$ reactions
at the threshold \cite{Ellis,Rekalo}.
It is found \cite{Rekalo} that
spin and parity conservation arguments result in a
precise prediction for the beam-target asymmetry $C^{BT}_{zz}$
for the $pp\to pp\Phi$ reaction
at the threshold,
\begin{equation}
C^{BT}_{zz}=\frac
{d\sigma(S_i=1) - d\sigma(S_i=0)}
{d\sigma(S_i=1) + d\sigma(S_i=0)} =1,
\label{as_thr}
\end{equation}
where $S_i$ is the total spin in the entrance channel. It is also
claimed \cite{Rekalo} that the decay matrix density amounts
$\rho_{00}=0, \rho_{11}=\rho_{-1-1} = \frac12$,
where the quantization axis is directed along the velocity of the
$\Phi$ meson. Real experiments, however, are performed at some finite energy
above threshold, therefore
the questions arise
(i) how are the threshold predictions modified
above threshold (say in the order of 100 MeV),  and
(ii) how sensitive are polarization observables to a certain
strangeness admixture in the proton wave function.
In this Letter we answer these questions. We show that the finite-energy
modification of the asymmetry is rather strong and that the strangeness
component modifies the previous predictions. We study
both the $pp\to pp\Phi$ and the $pn\to pn\Phi$ reactions.

We assume here that in the threshold-near region the strongest contributions
to the reaction come from the conventional
one-boson exchange mechanism and the $uud$ knock-out
by exchanged mesons, similar to the $\Phi$  photoproduction and
electroproduction \cite{Titov1,Henley}. (We do not consider here
the $s\bar s$ knock-out, thereby assuming that the corresponding
meson-$s\bar s$-$\Phi$ coupling constants are small.)
Both reaction channels are depicted in fig.~1 as Feynman diagrams.
We use the notation
$p_\alpha=(E_\alpha, {\bf p}_\alpha)$ with $\alpha=a,b,c,d$ for the
four-momenta of nucleons,
and $q=(E_\Phi, {\bf q})$) for the
produced $\Phi$ meson in the center of mass system (c.m.s.).
Hereafter $\vartheta$ denotes the polar $\Phi$ meson angle, and $s=E_a+E_b$.
We use a coordinate system with
${\bf z} \parallel {\bf p}_a$,
${\bf y} \parallel {\bf p}_a{\bf \times q}$.

{\bf 2. One-boson exchange model:}
Let us consider first the one-boson exchange (OBE) model for the conventional
$NN \to NN \Phi$ reaction dynamics.
Our OBE model is close to the previously employed
model \cite{Ko_PL}.
The meson-nucleon interaction Lagrangian reads
in obvious standard notation
\begin{equation}
{\cal L}_{MNN} =
-i g_{\pi NN} \bar N \gamma_5 \bbox{\tau} \bbox{\pi} N
- g_{\rho NN} \left( \bar N \gamma_\mu \bbox{\tau} N \bbox{\rho}^\mu
-\frac{\kappa_\rho}{2M_N} \bar N \sigma^{\mu\nu} \bbox{\tau}N
\partial_\nu\bbox{\rho}_\mu \right),
\label{L_MNN}
\end{equation}
while the $\Phi\rho\pi$ Lagrangian is
\begin{equation}
{\cal L}_{\Phi\rho\pi} = g_{\Phi\rho\pi} \, \epsilon^{\mu\nu\alpha\beta} \,
\partial_\mu \Phi_\nu \, {\rm Tr} ( \partial_\alpha \rho_\beta \pi),
\label{L_PRP}
\end{equation}
where  ${\rm Tr}(\rho \pi) = \rho^0 \pi^0 + \rho^+ \pi^- + \rho^- \pi^+$,
and bold face letters denote isovectors.
All coupling constants are dressed by monopole form factors
$F_i=(\Lambda^2-m_i^2)/(\Lambda^2-k_i^2)$, where $k_i$ is
the four-momentum of the exchanged meson.

The total invariant OBE amplitude is the sum of 4 amplitudes
\begin{equation}
T_{fi} = T_{fi}[ab;cd] + T_{fi}[ba;dc] +  
         T_{fi}[ab;dc] + T_{fi}[ba;cd],
\label{ampl1}
\end{equation}
where the last two terms stem from the antisymmetrization or
from charged meson exchange for $pp$ or $pn$ reactions, respectively.
The first term in eq.~(\ref{ampl1}) reads explicitly
\begin{equation}
T^r_{m_c,m_d;m_a,m_b}=
K(k_{\pi}^2,k_{\rho}^2)\,
\Pi_{m_d,m_b}(p_b,p_d)\,W_{m_c,m_a}(p_a,p_c)\,
\end{equation}
with
\begin{eqnarray}
K(k_{\pi}^2,k_{\rho}^2)
& = &
\frac{g_{\pi NN} \, g_{\rho NN} \, g_{\Phi\rho\pi}}
{(k^2_{\pi}-m_\pi^2)(k^2_{\rho}-m_\rho^2)}\,
\frac{\Lambda_\pi^2-m_\pi^2}{\Lambda_\pi^2-k^2_{\pi}}
\frac{\Lambda_\rho^2-m_\rho^2}{\Lambda_\rho^2-k^2_{\rho}},\nonumber\\[2mm]
\Pi_{m_d,m_b}(p_b,p_d)
& = &
\bar u(p_d,m_d)\gamma_5\, u(p_c,m_c),\nonumber\\[2mm]
W_{m_c,m_a}(p_a,p_c)
& = &
i\,\epsilon_{\mu\nu\alpha\beta}
k_{\rho}^\mu\Sigma^\nu_{m_c,m_a}
(k_{\rho})q_\Phi^\alpha \epsilon^{*\,r,\beta},
\label{ampl_not} \nonumber \\
\Sigma^\nu_{m_c,m_a}(k_{\rho})
& = &
\bar u(p_c,m_a)\left[\gamma^\nu - i \frac{\kappa_\rho}{2M_N}\sigma^{\nu\nu'}
{k_{\rho}}_{\nu'}\right]\, u(p_a,m_a).
\end{eqnarray}
$\epsilon^{r,\beta}$ denotes the $\Phi$ polarization four-vector,
and the index $r$ describes the polarization of the $\Phi$ meson
in the helicity basis; $m_a...m_d$ are the nucleons spin projection
on the quantization axis, and $k_{\rho}=p_c-p_a, \,k_{\pi}=p_b-p_d$.

The differential cross section is related
to the invariant amplitude $T_{fi}$ via
\begin{equation}
\frac{d^5\sigma}{dE_\Phi d\Omega_\Phi d\Omega_c}
=\frac{1}
{8\sqrt{s(s-4M_N^2)}(2\pi)^5}\,|T_{fi}|^2\,
\frac{qp_c^2}{|A p_c+C E_c|}
\end{equation}
where the energy $E_c$ of particle $c$
in the c.m.s. is defined by
\begin{equation}
E_c=\frac{AB\pm C\sqrt{B^2-M_N^2(A^2-C^2)}}{A^2-C^2},
\label{eq1}
\end{equation}
with
$A=(\sqrt{s}-E_\Phi),\,B=s-2E_\Phi\sqrt{s}+M_\Phi^2,\,
C=2q\cos \vartheta_{qp_c},$
where $\vartheta_{qp_c}$ is the angle between ${\bf q}$ and ${\bf p_c}$.
Both solutions in eq.~(\ref{eq1}) are to be taken into account.

{\bf 2.1 Fixing parameters:}
The coupling constant $g_{\Phi\to\rho\pi}$
is determined by the $\Phi\to\rho\pi$ decay.
Taking the most recent value $\Gamma( \Phi\to\rho\pi)$=0.69 MeV \cite{PDG}
we get $g_{\Phi\rho\pi}$ =1.10 GeV$^{-1}$.

The laboratory kinetic energy of the initial proton
in the region with $\Delta \equiv \sqrt{s} - \sqrt{s_0} \sim 0.1 \cdots 0.3$ GeV
is about  3 $\cdots$ 3.5 GeV
(here $\sqrt{s_0} = 2 M_N + M_\Phi$).
That is a quite large beam energy for the usual OBE model and, therefore,
one must use energy dependent coupling constants.
We use the
minimal energy dependence of ref.~\cite{Giessen}, where
each meson-nucleon coupling constant is modified by
a cut-off factor, i.e.
$g_{iNN} \to g_{iNN}^0 \, \mbox{exp} (-l_i \sqrt{s})$ with
$l_i = 0.11$ (0.18, 0.1, 0.1) GeV$^{-1}$ for $i = \pi$
($\rho, \sigma, \omega$).
We shall employ two sets of the OBE model parameters: set I relies on the
Bonn OBE model as listed in Table B.1 (Model II) of Ref.~\cite{Bonn},
and set II uses the results of ref.~\cite{Giessen}.
The cut-off parameter $\Lambda^\pi_{\Phi\rho\pi}$ is adjusted
by a comparison of additional calculations and
data \cite{pipnPhi} for the $\pi^- p\to n\Phi$
reaction. We find $\Lambda^\rho_{\Phi\rho\pi}=1.9$ GeV for set I,
and $\infty$ for set II.
These relatively large cut-off values are in agreement with the
pion photoproduction \cite{GG} and photon emission from the $V\pi\gamma$
($V=\rho,\,\omega$) vertices in $NN$ bremsstrahlung \cite{Jetter}.
Our cut-off parameter in set I is greater then the one found in
ref.~\cite{Ko_PL}
because we use energy-suppressed coupling constants and we
have a negative interference between vector and tensor parts
in the $pn\rho$ vertex which is intimately related to the definition
of the Lagrangian  eq.~(\ref{L_MNN}).
Note that we describe the data \cite{pipnPhi} in a limited region of
$\Delta_{\pi^- p \to n \Phi} \equiv \sqrt{s}-M_N+M_\Phi \le 0.15$ GeV.
At higher energies the OBE model overestimates the data. (In this region
one could use a stronger energy suppression in the vertices as
in ref.~\cite{Sibirtsev}.)
We have checked that the momenta at the $\pi \rho \Phi$ vertex
in the $NN \to NN \Phi$ reaction are in the
region where we describe the $\pi^- p \to n \Phi$ reaction correctly.
Taking into account the symmetry
of the off-shell mesons in the $\Phi\rho\pi$ vertex we use
$\Lambda^\pi_{\Phi\rho\pi} = \Lambda^\rho_{\Phi\rho\pi}$
for each parameter set.

{\bf 2.2 Threshold limit:}
At the threshold, where
$\Delta \to 0$,
one can neglect
terms proportional to $|{\bf q}|/M_\Phi$, $|{\bf p}_{c,d}|/M_N$
in the amplitude.
Meson propagators and form factors become constants
because they depend on the same variable
$k^2_{\rho,\pi}\to k_0^2 = - M_NM_\Phi$. Hence one can express
the amplitudes in the form
\begin{eqnarray}
T_{fi}^r[ab;cd]
& = &
T_0\,U_{fi}^r[ab;cd],\nonumber\\
T_0
& = &
K(k_0^2)\,(1-\kappa_\rho)M_NM_\Phi\sqrt{M_\Phi(M_N+\frac14 M_\Phi)},
\nonumber\\
U_{fi}^0[ab;cd]
& = &
2m_b\cos\vartheta_b\,
\delta_{m_a,-m_c}\delta_{m_b,m_d}\,\sin\vartheta, \nonumber\\
U_{fi}^{\pm}[ab;cd]
& = &
\mp \frac{1}{\sqrt{2}}\,2m_b\cos\vartheta_b\,
\delta_{m_a,-m_c}\delta_{m_b,m_d}\,(\cos\vartheta \pm 2m_a).
\label{tnr}
\end{eqnarray}
For $\vartheta=0$ the spin is transferred to the $\Phi$ meson
only at the $NN\rho$ vertex by a nucleon spin-flip. Then for the
initial polarization $m_a=\frac12$ ($-\frac12$)
only $T^+$  ($T^-$) is non-zero, while for $\vartheta=\pi$ only
$T^+$ ($T^-$) is non-zero for $m_a=-\frac12$ ($\frac12$).
Summation of $|T^r_{fi}|^2$ over spin projections of
the outgoing particles leads to the following expressions
for $pp$ and $pn$ collisions
\begin{eqnarray}
\sum_{m_{c,d}}|T^r_{fi,\alpha}|^2
& = &
T^2_0\, S^r_\alpha ,\nonumber\\
S^r_\alpha
& = &
\sum_{m_{c,d}}
\left(\xi_\alpha^1 U_{fi}^r[ab;cd] + \xi_\alpha^2 U_{fi}^r[ab;dc]
+\xi_\alpha^3  U_{fi}^r[ba;dc]  + \xi_\alpha^4 U_{fi}^r[ba;cd]\right)^2
\label{thr1}
\end{eqnarray}
with
$\xi^1_{pp}=\xi^3_{pp}=-\xi^2_{pp}=-\xi^4_{pp}=1$
$\xi^1_{pn}=\xi^3_{pn}=-1; \xi^2_{pn}=\xi^4_{pp}=2$.
Finally we get
\begin{equation}
S^0_{\alpha}=\Delta_{m_a,m_b}^\alpha \, \sin \vartheta,
\quad\quad
S^\pm_{\alpha}=\frac{1}{2} \Delta_{m_a,m_b}^\alpha \,
(\cos\vartheta \pm 2m_a)^2,
\label{thr2}
\end{equation}
where
$\Delta_{m_a,m_b}^{pp} = 4 (1+\delta_{m_a,m_b} - \delta_{m_a,-m_b})$
and
$\Delta_{m_a,m_b}^{pn} = 2 (1+4[1+\delta_{m_a,m_b} -\delta_{m_a,-m_b}])$.
Using these equations we get the threshold limits for the beam-target
asymmetry $C^{BT}_{zz} = 1$ and 0.8 for $pp$ and $pn$ collisions,
respectively. For both reactions the $\Phi$ spin density
matrix elements are $\rho_{00}=0$ and $\rho_{11}=\frac12$.
For the $pp$ reaction this result agrees with a previous
prediction \cite{Rekalo}.
Moreover, eqs.~(\ref{thr1}, \ref{thr2}) predict the ratio
of the corresponding total cross sections as
$\sigma_{np}^{tot}/ \sigma_{pp}^{tot}=80/(32\cdot \frac12)=5$,
where the additional factor $\frac12$ in the $pp$ cross section
represents the symmetry factor.

{\bf 3. $uud$ knock-out:}
The main ingredient of the knock-out photoproduction mechanism is the
assumption that the constituent quark wave function of the proton contains,
in addition to the usual 3-quark ($uud$) component,
a configuration with an explicit $s \bar s$ contribution.
A simple realization of this picture is the following wave function in the
Fock space \cite{Henley}
\begin{equation}
|p \rangle = A | [uud]^{1/2} \rangle
+
B \Bigl\{
a_0 | \bbox{[} [ uud]^{1/2} \otimes [s\bar s]^0 \bbox{]}^{1/2} \rangle
+
a_1 | \bbox{[} [ uud]^{1/2} \otimes [s\bar s]^1 \bbox{]}^{1/2} \rangle
\Bigr\},
\label{eq:wf}
\end{equation}
where $B^2$ denotes the strangeness admixture in the proton, and
$a_{0,1}^2 = \frac12$
are the fractions of the $s\bar s$ pair with spin 0 and 1, respectively.
The superscripts represent the spin of each cluster, and $\otimes$
represents the vector addition of spins of the $uud$ and $s\bar s$ clusters
and their relative orbital angular momentum ($\ell=1$).
Details on the wave functions in the relativistic harmonic oscillator
model \cite{RHOM} can be found in refs.~\cite{Titov1,Henley}.
We assume that the exchanged mesons interact with the
$uud$ cluster as with a structureless particle and describe
this interaction within the OBE model 
with exchanged $\pi,\,\rho,\,\omega,\,\sigma$ mesons, see fig.~1b.
The $s\bar s$ component is considered as spectator, that means only
the configuration with spin $S_{s\bar s}=1$ is realized.
The corresponding $S$ matrix element for the diagram with
meson exchange shown in fig.~1b reads
\begin{equation}
S_{fi}= \sum_{i=\pi,\rho,\omega,\sigma}
\left\langle c,\Phi\Bigl\vert \Gamma^i_\mu e^{ikx}
\Bigl\vert a\right\rangle\, D^{i\, \mu\nu}\,
\left\langle d\Bigl\vert \Gamma^i_\nu
\Bigl\vert b\right\rangle ,
\end{equation}
where $D^{\mu\nu}$ is the propagator and  $\Gamma_\nu$ is the
vertex function when the exchanged mesons are $\rho$ and $\omega$;
the Lorentz indices at the vertex and propagator
disappear for $\pi$ and $\sigma$ exchange.
The $T$ matrix is
calculated in the rest frame of the decaying nucleon and is expressed
via the two-body scattering $T$ matrix and the transition amplitude by
\begin{equation}
T_{fi} =
\sum_{i=\pi,\rho,\omega,\sigma} \sum_{m_x}
M^{N\to N\Phi}_{m_x,m_\Phi;m_a}\,
{}^iT^{NN\to NN}_{m_c m_d;m_x,m_b}(p_c,p_d;zp_a,p_b)
\end{equation}
($z$ is the momentum fraction carried by the $uud$ cluster)
with the abbreviations
\begin{eqnarray}
&&{}^iT^{NN\to NN}(pc,d;x,b)
=
i\bar u(c) \Gamma^i_\mu u(x)\, D^{i \, \mu\nu}
\bar u(d)  \Gamma^i_\nu u(b),
\nonumber \\
&&M^{N\to N\Phi}_{m_x,m_\Phi;m_a}
=
-i \sqrt{\frac32} \frac{V(q^*)}{ \gamma^*_\Phi \, \gamma^{*2}_c}
\sum_{m,J_c=0,1; m_c}
\left\langle 1 m_\Phi 1m\Bigl\vert J_c m_c \right\rangle\,
\left\langle J_c m_c \frac12 m_x
\Bigl\vert \frac12 m_a \right\rangle\,\hat q^*_m,
\end{eqnarray}
where ${\bf q}^*$ is the $\Phi$ momentum in the laboratory 
system, $\hat q^*_m$ denotes its projected unit vector in the circular basis,
and $V({\bf k})$ stands for the wave function of the relative motion
normalized as
$\int V^2({\bf k})d{\bf k}/(2\pi)^3/2\sqrt{({\bf k}^2+M_\Phi^2)}=1$;
$\gamma^*_{\Phi,c}$ is the corresponding Lorentz factor
which reflects the Lorentz contraction in the relativistic
constituent model. In our calculations we use a
Gaussian distribution $V(x) = N x \exp(-x^2/2\Omega)$ with
$\sqrt{\Omega}= 2.41$ fm$^{-1}$~\cite{Titov1,Henley}. The final amplitude
contains the sum over all exchanged mesons and consists
of 2 direct and 2 exchange amplitudes for the $pp$ reaction, and 2 direct
and 0 ($\omega, \sigma$) or 2 ($\pi, \rho$)
exchange amplitudes for the $pn$ reaction to be taken with their
proper isospin factors.

The corresponding amplitudes in the $\Phi$ helicity basis can be obtained
by
$  T^r = \sum_{m_\Phi} d^1_{m_\Phi,r} (\vartheta) \, T_{m_\Phi}$
with the Wigner rotation functions $d^1_{m_\Phi,r}$.
Note that this amplitude may be expressed in the covariant form
$T^r = W^\mu \, \epsilon^{r*}_\mu$, where the time component
$W^0$ is found from the transversality condition. The net result
is a renormalization of the component with $r=0$
as $T^0\to T^0 M_\Phi / E_\Phi$.

In the threshold limit the beam-target asymmetry is 1 ($pp$ reaction) or
$\sim 0$ ($pn$ reaction). $\sigma_{pn}^{tot}/\sigma_{pn}^{tot}$
depends on the model for the two-body $T$ matrix, and in our case
it is close to 1. The $\Phi$ spin density matrix elements here
coincide with the conventional OBE model predictions.

{\bf 4. Results:}
Our result for the total cross section in the $pp$ reaction is
shown in fig.~2.
The lower (upper) solid lines correspond to the conventional
OBE channel for the
set I with $\Lambda^{\rho,\pi}_{\Phi\rho\pi}=1.9$ GeV
($\Lambda^\rho_{\Phi\rho\pi}=1.9$ GeV,
 $\Lambda^\pi_{\Phi\rho\pi} = \infty$);
the short-dashed line is the prediction for set II.
The dashed line depicts the calculation with the constant matrix element of
eq.~(\ref{thr1}). One can see that up to
$\Delta \sim 1$ GeV the cross section is described fairly well
by the phase space integral alone. The space between upper solid and dashed
curves indicates upper and lower limits for the OBE model prediction.
In spite of the fact that the OBE model is reliable in the region of
$\Delta \le 0.2$ GeV it fits satisfactorily the available
experimental data \cite{Brenschede,Baldi} in a much wider interval.
The cross section for the $pn$ reaction is greater by a factor
$\simeq 5$ as given already in our threshold-near prediction above.
Open squares and black dots show the predictions
for the knock-out channel within the OBE model with sets I and II,
respectively,
and a strangeness probability $B^2=0.01$.
The distance between them is indicative for
the accuracy of the theoretical prediction.
Taking into account that, contrary to set II, the set I
overestimates the elastic $NN$ scattering at $-t\sim 1 \cdots 2$ GeV$^2$
by a factor of $20 \cdots 40$, the prediction based on set II seems
to be more realistic. So, we can conclude, that the  difference
between conventional OBE and knock-out channels for the total cross section
is about two orders of magnitude
at $\Delta \simeq 0.1$ GeV for $B^2=1\%$.
However, in the differential cross section this ratio changes, because
the knock-out model predicts a large enhancement for the
$\Phi$ production in forward and backward directions when the
recoil nucleons move along the $z$ axis. In this case
the difference between two channels drastically decreases
and becomes a factor $5\cdots8$.  The interference
between these two channels is negligible, and in coplanar
geometry it disappears because the OBE amplitude is real,
while the knockout amplitude is imaginary
since the exchanged meson is absorbed by the 5-quark component
in the proton wave function.

Figs.~3 and 4 show our predictions for the beam-target asymmetry
as a function of the $\Phi$ production angle $\vartheta$ at fixed recoil
nucleon angles at $\Delta = 0.1$ GeV and
$|{\bf q}| = \frac23 \lambda(s,M_\Phi^2,4M_N^2)/2\sqrt{s}$
for $pp$ and $pn$ reactions, respectively
($\lambda$ is the usual triangle function).
The left panels of figs.~3 and 4 show separately the asymmetry for pure
OBE and knock-out channels.
For the $pp$ reaction the long-dashed line is the threshold prediction.
One can see that at finite energy the asymmetry differs significantly
from the threshold value for both the conventional OBE (solid lines) and
knock-out (short-dashed line) mechanisms.
The same is seen for the $pn$ reaction,
where the dashed line is the threshold prediction for the knock-out
channel. The right panels of figs.~3 and 4 display
the asymmetry for the sum
of the two channels for two strangeness probabilities $B^2 =$ 2 and 5\%
(dashed and short-dashed curves).
One can see that the knock-out channel modifies
the OBE model prediction strongly in the
forward and backward directions in $pp$
reactions. In $pn$ reactions this modification is smaller because the
total cross section for the OBE channel is about 5 times greater
then in $pp$ reaction, while for the knock-out channel it is on same level.

The $\Phi$ meson spin density is modified too.
At finite  energy their modifications for the OBE channel are negligible
and the matrix elements are $\rho_{00}\simeq 0$, $\rho_{11}\simeq 0.5$.
The inclusion of knock-out channel changes them to
$\rho_{00} =$ 0.12 (0.22), $\rho_{11}\simeq$ 0.44 (0.39) for the
strangeness probabilities $B^2 =$ 2 (5)\%
in backward and forward directions.

{\bf 5. Summary:}
In summary we calculate within an extended OBE model
with a $uud$ knock-out mechanism the cross section and
polarization observables for the reaction $NN \to NN \Phi$.
While the total cross section is hardly sensitive to an admixture
of a $s \bar s$ configuration in the nucleon, a measurement of the
target-beam asymmetry should reveal the presence of hidden strangeness.
Already slightly above the threshold the interaction dynamics
becomes important and changes the threshold predictions.

{\bf Acknowledgments:}
Useful discussions with A. Brenschede are gratefully acknowledged.
One of the authors (A.I.T.) acknowledges the warm hospitality of the nuclear
theory group in the Research Center Rossenorf.
This work is supported by the BMBF grant 06DR829/1 and the
Heisenberg-Landau program.

\newpage

\vspace*{2cm}

\begin{figure}[h]
\centering
~\\[-1cm]
\centerline{\epsfxsize=.9 \hsize \epsffile{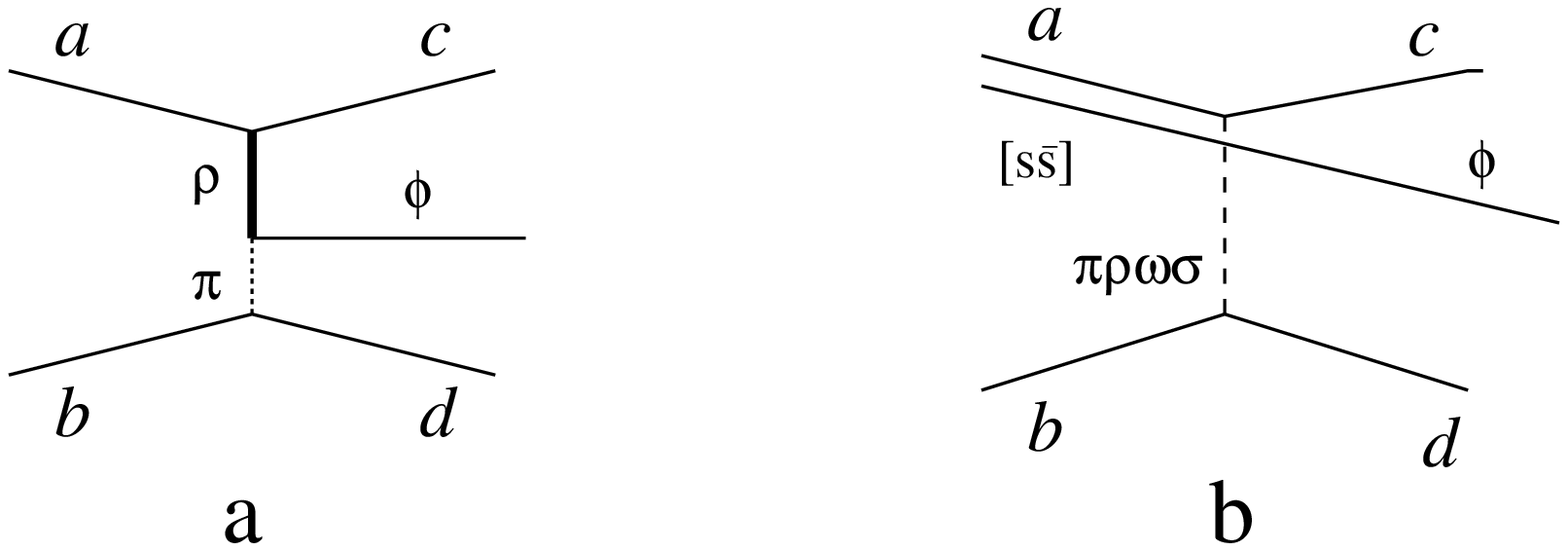}}
~\\[0.5cm]
\caption{
Tree level Feynman diagrams for the $\Phi$ production.
}
\label{fig1}
\end{figure}

\vfill

\begin{figure}[h]
\centering
~\\[.5cm]
\hspace*{-3cm}
\centerline{\epsfxsize=.8 \hsize \epsffile{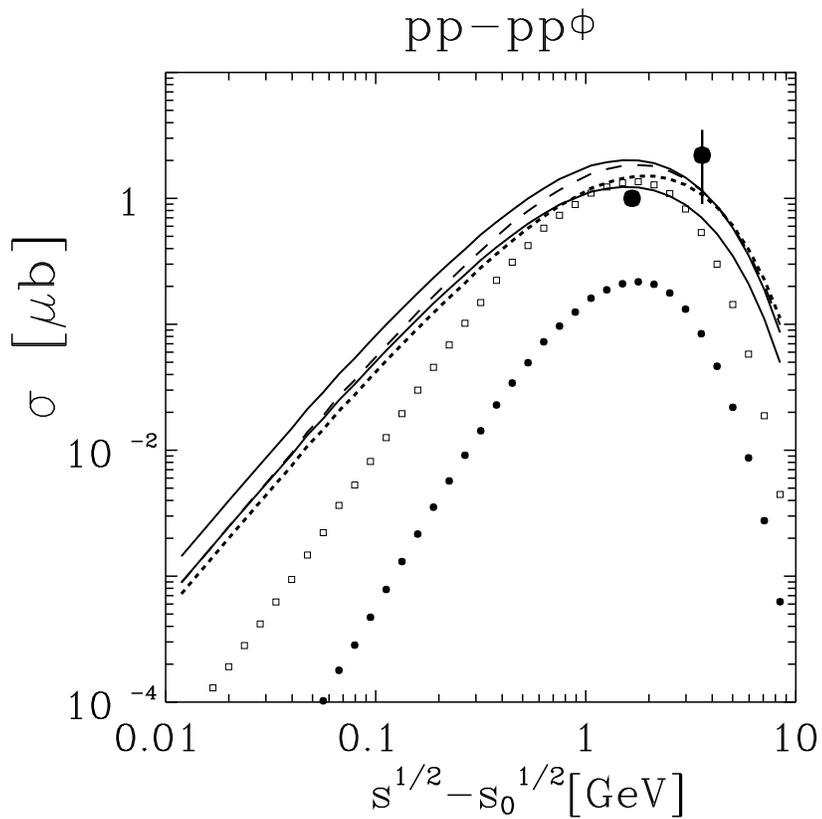}}
~\\[-3cm]
\caption{The total cross section fort the reaction $pp \to pp \Phi$.
The meaning of symbols and curves is described in the text.
Data from \protect\cite{Baldi}.
}
\label{fig2}
\end{figure}

\newpage

\begin{figure}[h]
\centering
~\\[7cm]
\hspace*{-4cm}
\centerline{\epsfxsize=.9 \hsize \epsffile{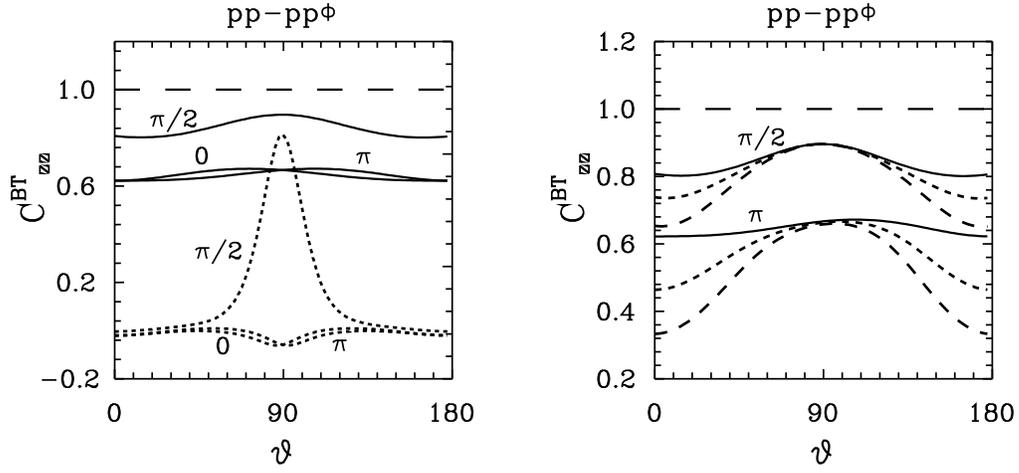}}
~\\[-4cm]
\caption{The beam-target asymmetry for the reaction $pp \to pp \Phi$
as a function of the c.m.s. polar angle $\vartheta$ of the
$\Phi$. Left (right) panel: contributions from conventional
OBE model and $uud$ knock-out mechanism separately (together).
The meaning of symbols and curves is described in the text.
}
\label{fig3}
\end{figure}

\vfill

\begin{figure}[h]
\centering
~\\[.1cm]
\hspace*{-4cm}
\centerline{\epsfxsize=.9 \hsize \epsffile{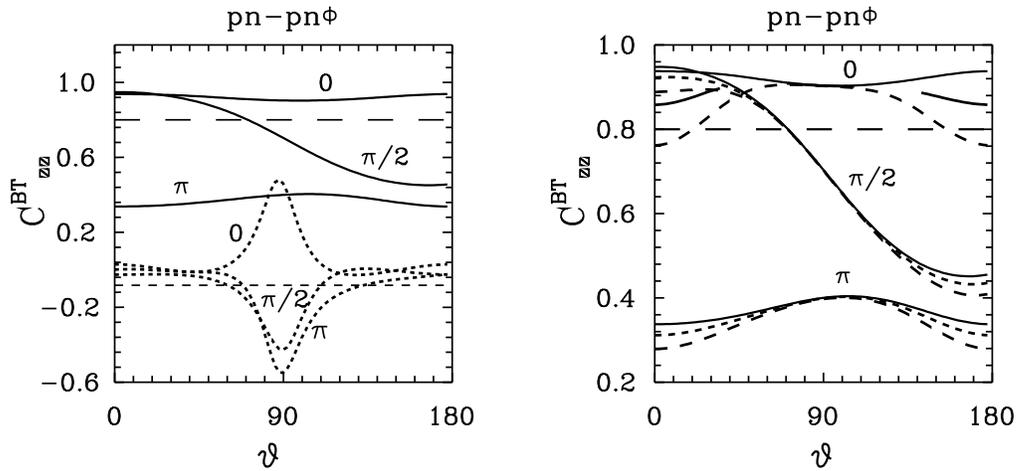}}
~\\[-4cm]
\caption{The same as in fig.~3 but for the reaction $pn \to pn \Phi$.
}
\label{fig4}
\end{figure}

\end{document}